# Ultra-Low-Energy Straintronics Using Multiferroic Composites

Kuntal Roy

School of Electrical and Computer Engineering, Purdue University,

West Lafayette, Indiana 47907, U.S.A.

## ABSTRACT

The primary impediment to continued improvement of charge-based electronics is the excessive energy dissipation incurred in switching a bit of information. With suitable choice of materials, devices made of multiferroic composites, i.e., strain-coupled piezoelectric-magnetostrictive heterostructures, dissipate miniscule amount of energy of ~1 attojoule at room-temperature, while switching in sub-nanosecond delay. Apart from devising memory bits, such devices can be also utilized for building logic, so that they can be deemed suitable for computing purposes as well. Here, we first review the current state of the art for building nanoelectronics using multiferroic composites. On a recent development, it is shown that these multiferroic straintronic devices can be also utilized for analog signal processing, with suitable choice of materials. By solving stochastic Landau-Lifshitz-Gilbert equation of magnetization dynamics at room-temperature, it is shown that we can achieve a voltage gain, i.e., these straintronic devices can act as voltage amplifiers.

## INTRODUCTION

Electric-field induced magnetization switching is a promising mechanism that can harness an energy-efficient binary switch replacing the traditional charge-based transistors for our future information processing systems [1,2]. With suitable choice of materials, when a voltage of few millivolts is applied across a multiferroic composite device, i.e., a magnetostrictive nanomagnet strain-coupled to a piezoelectric layer (see Fig. 1a) [3-5], it strains the piezoelectric layer and the generated strain is transferred to the magnetostrictive layer. Subsequently, a stress anisotropy is developed in the nanomagnet. This voltage-controlled magnetic anisotropy can switch the magnetization of the nanomagnet between its two stable states that store a binary information 0 or 1 [6-8]. This study has opened up a new field named straintronics [1,9,10] and experimental efforts to realize such devices are considerably emerging [11-14]. Although the experimental efforts have demonstrated the induced stress anisotropy in the magnetostrictive nanomagnets, the experimental demonstration of switching delay and utilizing low-thickness (< 25 nm) piezoelectric layers are still under investigation.

So far binary switching of the magnetization in magnetostrictive nanomagnets is investigated. While computing and signal processing tasks are indeed shifted to digital domain, sometimes *analog* signal processing is fundamentally necessary, e.g., processing of natural signals. When a transmitted signal is received at the receiver end, the signal is weak in magnitude and also noisy due to attenuation and noise in the environment. Hence, the signal needs to be amplified and filtered. Then the signal has to be converted to digital domain via analog-to-digital converter. Only thereafter, the signal is ready for processing in digital domain.

After digital signal processing, the signal again may need to be converted to analog domain by an digital-to-analog converter for transmission through environment. Hence, for different purposes, we do have the requirement of analog signal processing.

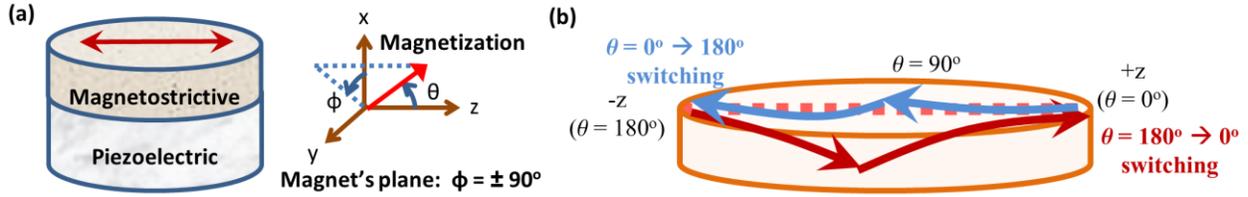

**Figure 1. (a)** Schematic diagram of a multiferroic composite, i.e., a strain-mediate piezoelectric-magnetostrictive heterostructure, and axis assignment. The mutually anti-parallel magnetization states of the single-domain magnetostrictive nanomagnet (shaped like an elliptical cylinder) along the z-axis store the binary information 0 or 1. In spherical coordinate system, $\theta$ is the polar angle and $\phi$ is the azimuthal angle. We will call the z-axis the easy axis, the y-axis the in-plane hard axis, and the x-axis the out-of-plane hard axis based on the chosen dimensions of the nanomagnet. **(b)** A complete 180° switching of magnetization is possible due to out-of-plane excursion (in x-direction) of magnetization [7]. While switching along the –y axis rather than +y axis (shown by arrows), the directions of the out-of-plane excursions would be opposite [7].

Here, we show that we can utilize multiferroic straintronic devices for ultra-low-energy analog signal processing purposes too. The applied voltage across the multiferroic composites generates stress on the magnetostrictive layer and this in turn modulates the potential landscape of the nanomagnet. The analog nature is harnessed by considering the fact that magnetization is not exactly stable at an easy axis rather it is fluctuating around a mean value due to thermal agitations. As the voltage modulates the potential landscape of the magnetostrictive nanomagnet, the mean value of magnetization gets shifted *continuously*. The measurement apparatus will measure this mean value of magnetization's orientation. Comparing this *continuous* mean value with respect to a hard (fixed) nanomagnet's magnetization, e.g., using tunneling magnetoresistance (TMR) measurement, a *continuous* output voltage while varying the input voltage can be produced. We solve stochastic Landau-Lifshitz-Gilbert (LLG) equation of magnetization dynamics at room-temperature [15-17] to demonstrate this concept of analog signal processing functionality.

**THEORY**

We model the magnetostrictive nanomagnet in the shape of an elliptical cylinder (see Fig. 1a). The total potential energy of the stressed polycrystalline single-domain nanomagnet is the sum of the shape anisotropy energy and the stress anisotropy energy [18]: $E = B(\phi) \sin^2\theta$, where $B(\phi) = B_{shape}(\phi) + B_{stress}$, $B_{shape}(\phi)$ is the $\phi$-dependent strength of shape anisotropy energy (proportional to $\cos^2\phi$ + constant term), which is minimum on magnet's plane $\phi = \pm 90°$ [1,7], and $B_{stress}$ is the strength of the stress anisotropy energy [1]. $B_{stress}$ is proportional to the magnetostriction coefficient $(3/2)\lambda_s$ and the stress $\sigma$. Materials like Terfenol-D has positive $(3/2)\lambda_s$, thus a compressive (i.e., negative) stress will favor alignment of magnetization along the minor axis (y-axis in Fig. 1a), and tensile (i.e., positive) stress along the major axis (z-axis).

The torque due to shape and stress anisotropy $\mathbf{T_E}$ is derived from the gradient of potential profile [1] and the torque due to thermal fluctuations $\mathbf{T_{TH}}$ is treated via a random magnetic field [1,17]. In the macrospin approximation, the magnetization $\mathbf{M}$ of the nanomagnet has a constant magnitude but a variable direction, so that we can represent it by a vector of unit norm $\mathbf{n_m}$ =$\mathbf{M}$/|$\mathbf{M}$|. The magnetization dynamics under the action of these two torques $\mathbf{T_E}$ and $\mathbf{T_{TH}}$ is described by the stochastic Landau-Lifshitz-Gilbert (LLG) equation [15-17] as follows:

$$\frac{d\mathbf{n_m}}{dt} - \alpha \left( \mathbf{n_m} \times \frac{d\mathbf{n_m}}{dt} \right) = -\frac{|\gamma|}{M}[\mathbf{T_E} + \mathbf{T_{TH}}],$$

where $\alpha$ is the phenomenological Gilbert damping parameter [16] through which magnetization relaxes towards the minimum energy position on magnet's potential landscape, $\gamma$ is the gyromagnetic ratio of electrons. Note that $\mathbf{M}$, $\mathbf{T_E}$, and $\mathbf{T_{TH}}$ are all proportional to the magnet's volume. Solving the above equation, we get the trajectory of magnetization over time.

**DISCUSSION**

We consider the magnetostrictive layer to be made of polycrystalline Terfenol-D, which has 30 times higher magnetostriction coefficient in magnitude than the common ferromagnetic materials (e.g., Fe, Co, Ni) and it has the following material properties - Young's modulus (Y): 80 GPa, magnetostrictive coefficient $(3/2)\lambda_s$: +900e-6, saturation magnetization ($M_s$): 8e5 A/m, and Gilbert damping parameter ($\alpha$): 0.1 [6]. We assume the following dimensions of the nanomagnet: 100 nm x 90 nm x 6 nm [6,19]. We do consider the distribution (rather than a fixed value) of initial orientation of magnetization due to thermal fluctuations, which has crucial consequence on device performance [6,10].

For the piezoelectric layer, we may use lead-zirconate-titanate (PZT), but using lead magnesium niobate-lead titanate (PMN-PT) is preferable since it has high $d_{31}$ (= 1300 p m/V) coefficient [20]. Considering that a maximum 500 ppm strain can be generated in the piezoelectric layer and about 75% strain gets transferred in the magnetostrictive nanomagnet, the maximum stress (Y. strain product) that can be produced would be 30 MPa. The piezoelectric layer is four times thicker than the magnetostrictive layer and hence 9.2 mVs of voltages would generate 30 MPa in the magnetostrictive Terfenol-D layer. Such miniscule amount of voltage is the basis of ultra-low-energy computing using these multiferroic devices [1,2]. Although, for simplicity, we have assumed only $d_{31}$ coefficient and uniaxial stress along the easy axis, we can consider also $d_{32}$ coefficient [14] which can allow us to work with even lower voltages.

**Digital memory and logic devices**

The usual perception is that stress can rotate magnetization only by 90° from ±z-axis to ±y-axis (see Fig. 1b), however, it is shown that out-of-plane excursion of magnetization can generate an intrinsic equivalent asymmetry, which can completely switch the magnetization by 180° [7]. Full 180° switching facilitates having the full tunneling magnetoresistance (TMR) while reading the magnetization state using a magnetic tunnel junction (MTJ) [7]. It should be noted that such switching leads to a toggle memory unless we have a mechanism to maintain the direction of switching [21]. The 90° switching mechanisms can be also utilized to direct switching in a particular direction, however, it gives us lower TMR.

While reading the magnetization state, a material issue crops up since magnetostrictive materials (Terfenol-D, iron-gallium) cannot be in general used as the free layer of an MTJ. Usually, CoFeB is used as the free layer alongwith MgO spacer that gives us high TMR [22]. Using half-metals can lead to even higher TMR [23]. This issue can be simply solved by introducing an insulator between the magnetostrictive nanomagnet and the free layer, and exploiting the magnetic dipole coupling in between to rotate them concomitantly [24].

Apart from using as memory bits, these straintronic multiferroic devices have been proposed for logic design purposes too [9,10]. It is shown that straintronic logic can operate using Bennett clocking mechanism in the presence of room-temperature thermal fluctuations [10]. However, a more intriguing mechanism is to utilize a single memory bit as universal logic gates [9]. Individual units can be concatenated since they have voltage gain and input-output isolation [9].

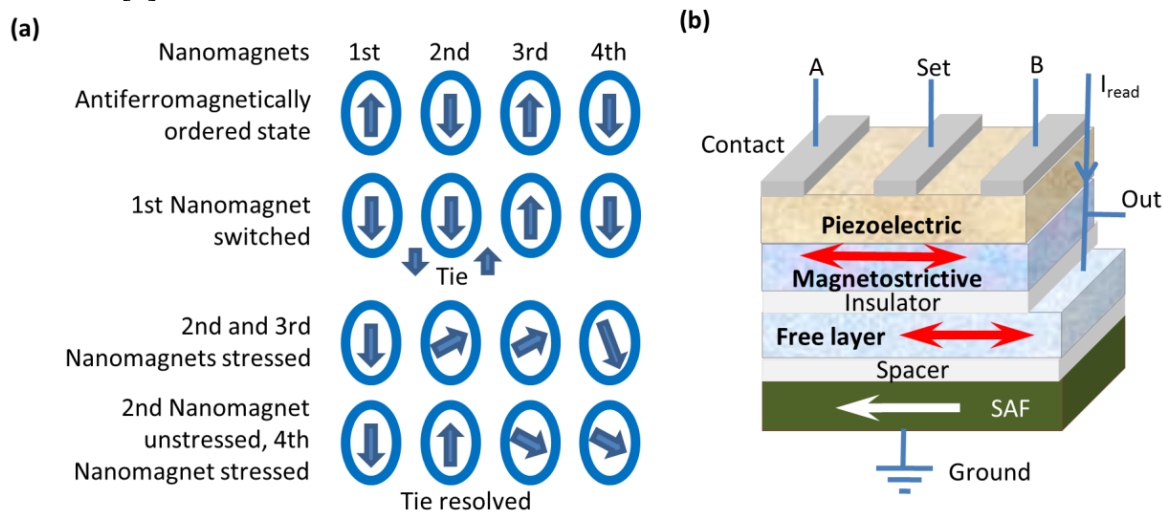

**Figure 2. (a)** Unidirectional information propagation through a horizontal chain of straintronic devices. Note that the dipole coupling between the neighboring nanomagnets is bidirectional and hence we need to impose the unidirectionality in time (using a 3-phase clocking scheme) [10]. (Reprinted with permission from Ref. 10. Copyright 2014, AIP Publishing LLC.) **(b)** Schematic diagram of single-element straintronic universal logic gates [9]. By applying voltages at the terminals A and B, the magnetization of the magnetostrictive nanomagnet can be switched.

## Analog signal processing

The potential landscape of a single-domain nanomagnet is a bistable double-well separated by an energy barrier, which is suitable for digital signal processing. For digital switching and information storage, the potential energy barrier is removed by an external input and there should be a force (externally applied or internal) to favor the final state of the magnetization. Then the barrier is restored to complete the switching cycle and the magnetization stays in its final state, so that non-volatile operation is achieved. However, for analog signal processing, we need to stop abrupt switching between the two states and it requires a *continuous* rotation of magnetization while varying the input voltage. We can make the potential landscape monostable by utilizing dipole coupling from a neighboring nanomagnet. By solving stochastic

LLG equation in the presence of room-temperature thermal fluctuations we show that it is possible to get a *continuous* rotation of magnetization while varying the input voltage/stress (Fig. 3) and we can get a high-gain region in the input-output characteristics of such devices (Fig. 4).

An AC voltage of frequency 1 GHz is applied at the input terminal of the device (see Fig. 4a) and the output voltage is determined by first solving the stochastic LLG equation for magnetization dynamics and then getting the TMR measurement of the MTJ [22]. A voltage gain of 50 is achieved, while expending a miniscule amount of energy of less than 1 attojoule/cycle at room-temperature [25]. This amplifier also filters out the high-frequency noise [25]. A rectifier characteristics can be obtained from these analog straintronic devices too. The degree of rectification capability increases with the TMR of the MTJ [25].

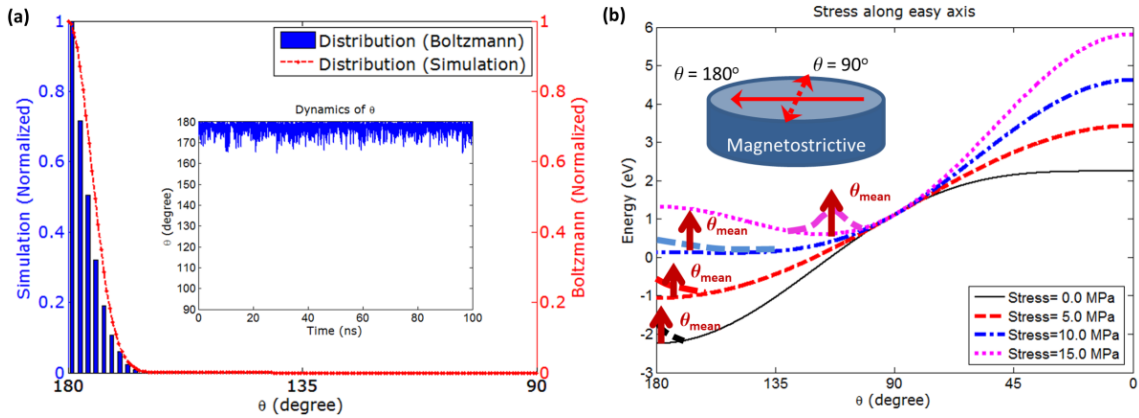

**Figure 3.** (a) Magnetization is fluctuating due to room-temperature (300 K) thermal fluctuations around one easy axis $\theta=180^o$. This is a Boltzmann distribution. (b) As we modulate the potential landscape of the magnetostrictive nanomagnet with stress, the energy barrier decreases and the mean orientation of the magnetization changes *gradually* rather than very abruptly.

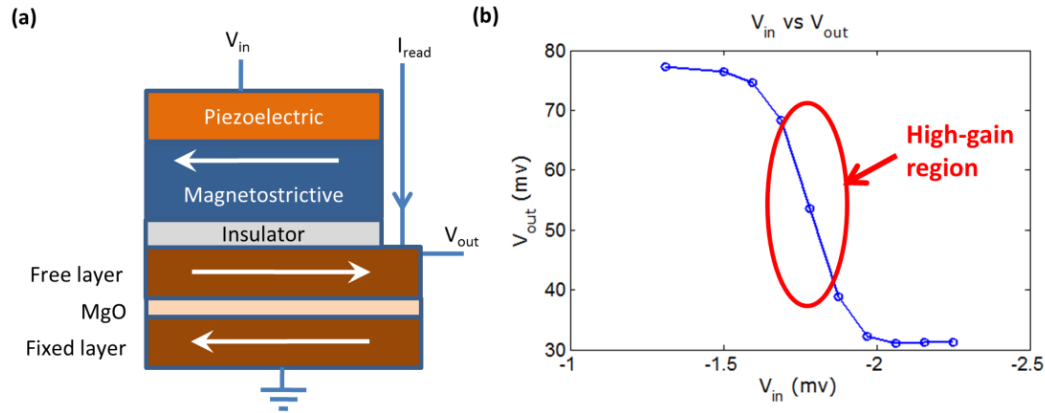

**Figure 4.** (a) The device structure to harness the analog nature consists of a multiferroic composite attached to an MTJ separated by an insulator. The magnetostrictive nanomagnet is magnetically coupled (but electrically separated) to the free layer via dipole coupling. $I_{read}$ is a constant read current and the output voltage $V_{out}$ varies *continuously* with $V_{in}$. (b) The relation between the input voltage $V_{in}$ and the output voltage $V_{out}$. This is an inverter like characteristics with a high-gain transition region. If we bias the input voltage in this high-gain region, we would get an amplified output voltage.

# CONCLUSIONS

Multiferroic devices hold profound promise for energy-efficient computing in beyond Moore's law era. With suitable choice of materials, these devices can perform both digital and analog computing tasks. Intrinsic coupling between the polarization and magnetization can facilitate complete 180$^o$ switching apart from utilizing the out-of-plane excursion of magnetization. By optimizing the device dimensions e.g., increasing thickness while decreasing the lateral dimensions can facilitate scaling too. Due to the miniscule energy dissipation, these devices can be operated by energy harvested from the environment. This is an unprecedented opportunity in ultra-low-energy computing that can perpetuate Moore's law to beyond the year 2020. Successful experimental implementation can pave the way for our future nanoelectronics.

# ACKNOWLEDGMENTS

This work was supported in part by FAME, one of six centers of STARnet, a Semiconductor Research Corporation program sponsored by MARCO and DARPA.

# REFERENCES

1. K. Roy, *SPIN* **3**, 1330003 (2013).
2. K. Roy et al, *Appl. Phys. Lett*. **99**, 063108 (2011). News: *Nature* **476**, 375 (2011).
3. N. A. Spaldin and M. Fiebig, *Science* **309**, 391 (2005).
4. W. Eerenstein et al, *Nature* **442**, 759 (2006).
5. N. A. Pertsev, *Phys. Rev. B* **78**, 212102 (2008).
6. K. Roy et al, *J. Appl. Phys.* **112**, 023914 (2012).
7. K. Roy et al, *Sci. Rep.* **3**, 3038 (2013).
8. K. Roy et al, *Phys. Rev. B* **83**, 224412 (2011).
9. K. Roy, *Appl. Phys. Lett.* **103**, 173110 (2013).
10. K. Roy, *Appl. Phys. Lett.* **104**, 013103 (2014).
11. N. Tiercelin et al, *Appl. Phys. Lett.* **99**, 192507 (2011).
12. N. Lei et al, *Nature Commun.* **4**, 1378 (2013).
13. H. K. D. Kim et al, *Nano Lett*. **13**, 884 (2013).
14. T. Jin et al, *Appl. Phys. Express* **7**, 043002 (2014).
15. L. Landau and E. Lifshitz, *Phys. Z. Sowjet.* **8**, 101 (1935).
16. T. L. Gilbert, *IEEE Trans. Magn.* **40**, 3443 (2004).
17. W. F. Brown, *Phys. Rev.* **130**, 1677 (1963).
18. S. Chikazumi, *Physics of Magnetism* (Wiley New York, 1964).
19. M. Beleggia et al, *J. Phys. D: Appl. Phys.* **38**, 3333 (2005).
20. Y. Jia et al, *Appl. Phys. Lett.* **88**, 242902 (2006).
21. M. Fechner et al, *Phys. Rev. Lett.* **108**, 197206 (2012).
22. S. S. P. Parkin et al, *Nature Mater.* **3**, 862 (2004).
23. T. Graf et al, *IEEE Trans. Magn.* **47**, 367 (2011).
24. K. Roy, unpublished.
25. K. Roy, unpublished.